\documentclass[twocolumn,showpacs,aps,prd,superscriptaddress]{revtex4}
\usepackage{graphicx}
\usepackage{dcolumn}
\usepackage{epsfig}

\RequirePackage{xspace}

\usepackage{relsize}
\def\babar{\mbox{\slshape B\kern-0.1em{\smaller A}\kern-0.1em
    B\kern-0.1em{\smaller A\kern-0.2em R}}}

\def\Kbar  {\kern 0.2em\overline{\kern -0.2em K}{}\xspace}

\def\Kz    {\ensuremath{K^0}\xspace}
\def\Kzb   {\ensuremath{\Kbar^0}\xspace}
\def\KzKzb {\ensuremath{\Kz \kern -0.16em \Kzb}\xspace}
\def\Kp    {\ensuremath{K^+}\xspace}
\def\Km    {\ensuremath{K^-}\xspace}

\def\KpKm  {\ensuremath{\Kp \kern -0.16em \Km}\xspace}

\def\Dbar    {\kern 0.2em\overline{\kern -0.2em D}{}\xspace}

\def\Dz      {\ensuremath{D^0}\xspace}
\def\Dzb     {\ensuremath{\Dbar^0}\xspace}
\def\DzDzb   {\ensuremath{\Dz {\kern -0.16em \Dzb}}\xspace}
\def\Dp      {\ensuremath{D^+}\xspace}
\def\Dm      {\ensuremath{D^-}\xspace}

\def\DpDm    {\ensuremath{\Dp {\kern -0.16em \Dm}}\xspace}

\def\Bbar    {\kern 0.18em\overline{\kern -0.18em B}{}\xspace}

\def\BB      {\ensuremath{B\Bbar}\xspace} 
\def\Bz      {\ensuremath{B^0}\xspace}
\def\Bzb     {\ensuremath{\Bbar^0}\xspace}
\def\BzBzb   {\ensuremath{\Bz {\kern -0.16em \Bzb}}\xspace}
\def\Bu      {\ensuremath{B^+}\xspace}
\def\Bub     {\ensuremath{B^-}\xspace}

\def\BpBm    {\ensuremath{\Bu {\kern -0.16em \Bub}}\xspace}

\mathchardef\Upsilon="7107
\def\Y#1S{\ensuremath{\Upsilon{(#1S)}}\xspace}

\def\FourS {\Y4S}

\mathchardef\Deltares="7101
\mathchardef\Xi="7104
\mathchardef\Lambda="7103
\mathchardef\Sigma="7106
\mathchardef\Omega="710A

\def\Deltabar{\kern 0.25em\overline{\kern -0.25em \Deltares}{}\xspace}
\def\Lbar{\kern 0.2em\overline{\kern -0.2em\Lambda\kern 0.05em}\kern-0.05em{}\xspace}
\def\Sigbar{\kern 0.2em\overline{\kern -0.2em \Sigma}{}\xspace}
\def\Xibar{\kern 0.2em\overline{\kern -0.2em \Xi}{}\xspace}
\def\Obar{\kern 0.2em\overline{\kern -0.2em \Omega}{}\xspace}
\def\Nbar{\kern 0.2em\overline{\kern -0.2em N}{}\xspace}
\def\Xb{\kern 0.2em\overline{\kern -0.2em X}{}\xspace}

\newcommand{\tev}{\ensuremath{\mathrm{\,Te\kern -0.1em V}}\xspace}
\newcommand{\gev}{\ensuremath{\mathrm{\,Ge\kern -0.1em V}}\xspace}
\newcommand{\mev}{\ensuremath{\mathrm{\,Me\kern -0.1em V}}\xspace}
\newcommand{\kev}{\ensuremath{\mathrm{\,ke\kern -0.1em V}}\xspace}
\newcommand{\ev}{\ensuremath{\mathrm{\,e\kern -0.1em V}}\xspace}
\newcommand{\gevc}{\ensuremath{{\mathrm{\,Ge\kern -0.1em V\!/}c}}\xspace}
\newcommand{\mevc}{\ensuremath{{\mathrm{\,Me\kern -0.1em V\!/}c}}\xspace}
\newcommand{\gevcc}{\ensuremath{{\mathrm{\,Ge\kern -0.1em V\!/}c^2}}\xspace}
\newcommand{\mevcc}{\ensuremath{{\mathrm{\,Me\kern -0.1em V\!/}c^2}}\xspace}

\def\mus  {\ensuremath{\rm \,\mus}\xspace}

\def\mus        {\ensuremath{\,\mu{\rm s}}\xspace}    

\def\to                 {\ensuremath{\rightarrow}\xspace}

\def\pep2{PEP-II}

\newcommand{\dedx}{\ensuremath{\mathrm{d}\hspace{-0.1em}E/\mathrm{d}x}\xspace}

\def\gsim{{~\raise.15em\hbox{$>$}\kern-.85em
          \lower.35em\hbox{$\sim$}~}\xspace}
\def\lsim{{~\raise.15em\hbox{$<$}\kern-.85em
          \lower.35em\hbox{$\sim$}~}\xspace}

\def\eps{\varepsilon\xspace}

\xspace

\newcommand{\jprlBase}       {Phys.\ Rev.\ Lett.\xspace}
\newcommand{\jprBase}        {Phys.\ Rev.\xspace}
\newcommand{\jplBase}        {Phys.\ Lett.\xspace}

\newcommand{\nimBaseC}       {Nucl.\ Instr.\ and Methods\xspace}

\newcommand{\zpBase}         {Z.\ Phys.\xspace}

\newcommand{\nima}      [1]  {\nimBaseC~A~{\bf #1}}

\newcommand{\plb}       [1]  {\jplBase\ B~{\bf #1}}

\newcommand{\jprl}      [1]  {\jprlBase\ {\bf #1}}
\newcommand{\jprd}      [1]  {\jprBase\ D~{\bf #1}}

\newcommand{\zpc}       [1]  {\zpBase\ C~{\bf #1}}

\def\jetset74   {\mbox{\tt Jetset \hspace{-0.5em}7.\hspace{-0.2em}4}\xspace}

\def\Bu{\mbox{$\mathrm{B_u}$}}

\def\Dz{\mbox{$\mathrm{ D^0}$}}

\def\plm{$\pm$}

\long\def\inst#1{\par\nobreak\kern 4pt\nobreak
    {\it #1}\par\vskip 10pt plus 3pt minus 3pt}
\begin{document}

\title{\boldmath Measurement of the Average $\phi$ Multiplicity in $B$ Meson Decay}

%
\author{B.~Aubert}
\author{R.~Barate}
\author{D.~Boutigny}
\author{J.-M.~Gaillard}
\author{A.~Hicheur}
\author{Y.~Karyotakis}
\author{J.~P.~Lees}
\author{P.~Robbe}
\author{V.~Tisserand}
\author{A.~Zghiche}
\affiliation{Laboratoire de Physique des Particules, F-74941 Annecy-le-Vieux, France }
% ...
\collaboration{The \babar\ Collaboration}
\noaffiliation
\date{\today}

\begin{abstract}
We present a measurement of the average multiplicity of $\phi$ mesons in $B^0$,
$\kern 0.18em\overline{\kern -0.18em B}{}^0$ and
$B^\pm$ meson decays. Using $17.6~\mathrm{fb}^{-1}$ of data taken at the
$\Upsilon{(4S)}\xspace$ resonance by the
\mbox{\slshape B\kern-0.1em{\smaller A}\kern-0.1em B\kern-0.1em{\smaller A\kern-0.2em R}}
detector at the PEP-II $e^+e^-\xspace$ storage ring at the Stanford 
Linear Accelerator Center, we reconstruct $\phi$ mesons in the $K^+K^-$ decay mode
and measure ${\cal{B}}(B\rightarrow \phi X) = (3.41\pm0.06\pm0.12)\%$.
This is significantly more precise than any previous measurement.
\end{abstract}

\maketitle

\section{Introduction}
The large data sample collected by the \babar\ detector provides an 
excellent opportunity for a significant improvement to the existing measurements of the
average $\phi$ multiplicity in $B$ meson decay at the \FourS resonance.
This quantity, which is conventionally denoted ${\cal{B}}(B\to\phi X)$,
was previously measured by CLEO as (2.3\plm0.6\plm0.5)\% \cite{CLEO}
and by ARGUS as (3.90\plm0.30\plm0.35)\% \cite{ARGUS}. These two measurements
disagree at the $1.8\sigma$ level, leading to a large error on the
Particle Data Group average ($3.5\pm0.7$)\% \cite{PDG}.
The OPAL collaboration has measured the average multiplicity 
${\cal B}(b\rightarrow \phi X)=(2.82\pm0.13\pm0.19)\%$ \cite{OPAL} at the $Z^0$ pole. 
This latter measurement is sensitive to 
$b$-hadron decays that are not accessible at \FourS experiments, including $b$
baryons and, in particular, the $B_s^0$ meson.

An improved measurement of ${\cal B}(B\to\phi X)$ can lead to improved measurements of
the $B_s^0$ oscillation frequency. The primary decay modes
of the $B_s^0$ meson contain $D_s^+$ mesons, which often (18\% \cite{DstoPhi})
produce a $\phi$ meson in their decays. Due to this high rate, $B_s^0$ decays into $\phi$ 
mesons are a prime decay chain for $B_s^0$ oscillation searches. An important
input to such searches is the knowledge of the background arising from 
non-strange $B$ meson decays into $\phi$ mesons.

Given the large size of the \babar\ data sample, this measurement is limited
by systematic errors. As a result, this analysis
is designed to minimize these systematic errors.  Minimal selection
criteria are applied, and efficiencies and backgrounds are evaluated
directly from data where possible.
The measurement is performed in $\phi$-momentum intervals
to minimize the systematic effects that may be introduced by differences
between the $\phi$ momentum spectrum in data and simulation.

\section{The \babar\ detector and data samples}
\label{detector}
The data used in this analysis were collected by the \babar\ detector at the \pep2
storage ring.
We use $17.6~\mathrm{fb}^{-1}$ of data taken at the \FourS resonance (on-resonance) 
and $4.1~\mathrm{fb}^{-1}$ of data taken at a center-of-mass energy
$20\mev$ below the \BB threshold
(off-resonance). The latter sample is used for the subtraction of the non-\BB component 
(continuum) in the on-resonance data. These data samples were taken between January and May 2002. 
Additional data, consisting of 3.5~$\mathrm{fb}^{-1}$ of on-resonance data and 
1 $\mathrm{fb}^{-1}$ of off-resonance data taken under different running conditions,
are used for verification of the result. 

A detailed description of the \babar\ detector is presented in
Ref.~\cite{ref:babar}. The components of the detector most
relevant to this analysis are described here.
Charged-particle tracks are reconstructed with a five-layer, double-sided
silicon vertex tracker (SVT) and a 40-layer drift chamber (DCH) with a
helium-based gas mixture, placed in a 1.5-T solenoidal field produced
by a superconducting magnet. The resolution on $p_T$, the charged-track
momentum transverse to the beam direction, is approximately
$(\delta p_T/p_T)^2 = (0.0013\ (\mbox{GeV}/c)^{-1}\, p_T)^2 + (0.0045)^2$.
Charged particles are identified from the ionization energy loss
(\dedx) measured in the DCH and SVT, and the Cherenkov radiation
detected in a ring-imaging Cherenkov device. The efficiency for
identifying true kaons exceeds 80\% over most of the momentum range of
interest, while the probability for a pion to be misidentified as a
kaon is less than 3\%.

We use Monte Carlo samples of $\Y4S \to \BzBzb$ and 
$B^+B^-$ decays, corresponding to twice the expected 
number of $B$ mesons in the data sample, to study our selection efficiency.
The $B$-meson decays are simulated according to previously measured branching fractions which
account for approximately 60\% of all $B$ decays. The remaining 40\%
are modeled by JETSET \cite{jetset}, while preventing any enhancement of
the first 60\%. The detector response in these samples is simulated with the GEANT4 
program \cite{geant} and cross-checked with control samples in the data.

\section{Event and candidate selection}
Events are selected if at least three tracks are found 
and the measured total energy is at least 4.5 \gev.
In order to suppress continuum background, events are rejected 
if the ratio of the second-to-zeroth order Fox Wolfram moments \cite{FW}
($R_2$) is higher than 0.25. This requirement 
rejects 62\% of the off-resonance data, while retaining 78\% of the 
simulated $B\to\phi X$ events.

The selection of $\phi\to K^+K^-$ candidates requires two
oppositely-charged tracks that satisfy
$0.1 < p_T < 10 \gevc$, have at least 12 hits in the DCH,
are consistent with originating from the primary interaction
point, and satisfy kaon identification criteria based on \dedx 
measurements and Cherenkov radiation.
Tracks are assigned a kaon mass hypothesis and neutral two-track combinations are formed.
Candidates are selected if their invariant mass is in the range 
$1.004<m_{KK}<1.036~\gevcc$. 
This mass window is equivalent to about 4.5 standard deviations
on either side of the nominal $\phi$ mass, where the RMS spread in 
the $m_{KK}$ distribution is 
due to both the natural $\phi$ width and the detector resolution. 
This relatively large
acceptance is chosen to reduce the effect of a possible mass resolution 
difference between data and Monte Carlo, at the expense of signal-to-background 
significance.

A total of $471,941$ ($34,900$) $\phi$ candidates survive these selection 
criteria in the on- (off-)resonance sample.
 
\section{Background estimation}
\label{section:bg}
Two sources of background to the $B\to\phi X$ signal are considered: 
random combinations of tracks that pass the selection (combinatorial background)
and true $\phi$ mesons that do not originate from \BB events
(continuum background). Because the reconstruction efficiency depends on the
momentum of the $\phi$, these backgrounds are subtracted separately in 16
bins of $\phi$ momentum.

We first remove the continuum background from our signal by subtracting the $m_{KK}$
distribution obtained in the off-resonance sample from that in 
the on-resonance sample, scaled by the ratio of the luminosities of the two samples. 
This scale factor is calculated by comparing the number of $e^+e^-\to \mu^+\mu^-$
events in the two samples.
The center-of-mass momenta of $\phi$ candidates in the off-resonance data
are scaled by the ratio of on/off-resonance beam energies to
account for the slightly different momentum spectrum of the continuum component
in the on-resonance sample.
This procedure explicitly accounts for all backgrounds from physics processes
other than \FourS production as their cross sections are almost identical
at the two energies; it also accounts for beam-related backgrounds, as the running
conditions were very similar.

We next subtract the combinatorial background to extract the number of $\phi$ mesons.
This background is estimated by fitting the mass distribution in sideband regions well away from both the
signal and the $KK$ threshold. The ranges $0.989<m_{KK}<1.002~\gevcc$ and $1.04<m_{KK}<1.1~\gevcc$
were chosen and the function $(m_{KK}-2m_K)^a\cdot(b+c\cdot m_{KK}+d\cdot m_{KK}^2+e\cdot m_{KK}^3)$
was used. This function provides a good description of the phase space in the vicinity of a 
threshold; a fit to the combinatoric $m_{KK}$ spectrum (removing true $\phi$ mesons) in Monte Carlo
gives $\chi^2=104$ for 95 degrees of freedom.

\begin{figure}
\begin{center}
\includegraphics[width=3.5in]{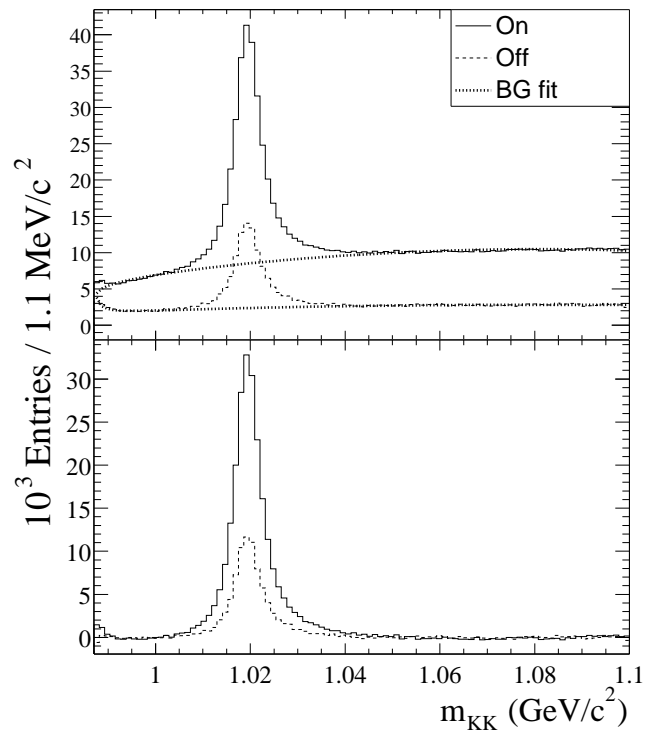}
\end{center}
\caption{Top: Invariant mass distributions of candidates passing all selection requirements
except that for the mass. The solid histogram shows candidates in the on-resonance
data sample, while the dashed histogram shows the off-resonance sample, scaled to the
luminosity of the on-resonance data. The fitted combinatorial background is overlaid
on both histograms as a dotted curve. Bottom: Resulting signal after
combinatorial background subtraction. Again, the solid histogram
represents on-resonance candidates, and the dashed histogram represents the
luminosity-scaled off-resonance candidates. Note that the vertical scale is displaced from
zero. The bump visible in the lowest bins of these plots is a threshold effect, which
is well understood, and the fit range was chosen so that this does not contribute
to the measurement.}
\label{OnOff}
\end{figure}

Figure~\ref{OnOff} shows the $m_{KK}$ distributions of the on-resonance and luminosity-scaled
off-resonance data samples for all $\phi$ momentum bins combined;
the fitted combinatorial background shapes are overlaid. The
on- (off-) resonance fit has $\chi^2=86\ (75)$ for 61 degrees of freedom.
With estimates of the combinatorial
background shape, the signal is extracted by subtracting the background.
The resulting signals are shown in the lower part of Figure~\ref{OnOff}. In the on-resonance
sample, we observe $(2.349\pm 0.007)\times 10^5\ \phi$ candidates, and the corresponding number
in the off-resonance sample is $(1.95\pm 0.02)\times 10^4$. These plots and numbers are representative only;
they are not used in the signal extraction.

Erroneous estimation of the signal in data and simulation could arise from
a number of sources. The phase-space function may be unable to
describe the background shape correctly (see Monte Carlo fit above). There
may be $\phi$ mesons with a reconstructed mass in the sideband regions,
which would affect the background estimation and therefore the signal
yield. The Monte Carlo may not correctly model the signal lineshape
(in particular, the fraction of $\phi$ mesons outside the signal mass region),
leading to a systematic uncertainty on the efficiency.
We consider each of these sources below.

We vary the fitting procedure in a number of ways in order to test the 
robustness of our background estimation.
We replace the third order polynomial in the above function with both a 
second order polynomial in $m_{KK}$ and an exponential term $\exp (b\cdot m_{KK})$.
These changes result in a 0.15\% and 0.65\% change in the number of
$\phi$ candidates respectively.

To account for the possibility that the reconstructed $\phi$ mass extends into the sideband regions,
we vary the upper bound of the region excluded from the fit, raising it from 1.04 to 
1.06 $\gevcc$ (while keeping the signal region as defined above).
The largest difference in the number of $\phi$ mesons in the signal region in
the above variations is 2.4\%.

Finally, we look at the fraction of candidates, after background subtraction,
outside the signal mass region. We count the number of candidates in the range 
$1.036<m_{KK}<1.05~\gevcc$ and calculate the ratio of this number to
the number of candidates in the signal region. This ratio is found to be 2.4\%
in data and 2.6\% in Monte Carlo. This yields a difference of 0.2\% on the number of
$\phi$ candidates between data and Monte Carlo.

We take the largest difference in all the above tests (2.4\%) to be 
the systematic uncertainty associated with the combinatorial background 
subtraction and signal selection.

\section{Selection efficiency}
\label{sec:recoeff}
\begin{figure}[h]
\begin{center}
\includegraphics[width=3.5in]{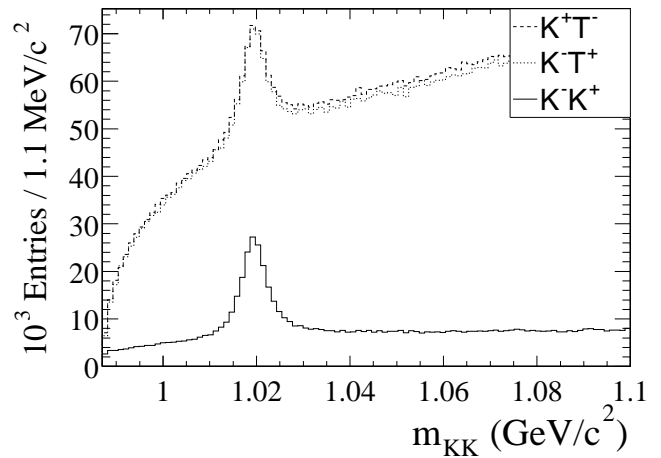}
   \caption{Mass distribution of the $K^+K^-$, $K^+ T^-$ and 
$K^- T^+$ samples. All selection criteria were applied in the $K^+K^-$ sample but one
track is not required to pass the kaon selection in the  $K^\pm T^\mp$ samples.}
   \label{fig:PID}
\end{center}
\end{figure}

The kaon identification efficiency is extracted from data to avoid the
systematic errors associated with Monte Carlo-based determinations. To 
do this, $\phi$ candidates are constructed from two-track 
combinations with at least one track passing the kaon identification criteria.
This is done for positive and negative tracks separately to account 
for a possible asymmetry. Three subsamples of the data are defined: $K^+K^-$, where
both tracks have passed the kaon identification requirements, 
and $K^\pm T^\mp$, where only one track
is required to pass the kaon selection.
The same off-resonance subtraction and $R_2$ requirement
is made in defining these samples as for the standard selection. The kaon 
identification efficiency is then given by the ratio 
of the number of $\phi$ mesons reconstructed in the $K^+K^-$ sample to the number
in the $K^\pm T^\mp$ samples: $\eps_{K^\pm}=N\phi_{KK}/N\phi_{K^\mp T^\pm}$.
Figure \ref{fig:PID} shows the invariant mass distribution of the
$K^+K^-$, $K^+ T^-$ and $K^- T^+$ samples over the entire momentum region.

Studies of data and Monte Carlo samples show that the kaon identification
efficiency is not constant throughout the
momentum range of our $\phi$ sample, but can be described by one constant 
efficiency values below $p_\phi=1.2~\gevc$ and another above this value. This step in efficiency is
caused by the transition from \dedx -based particle identification at low momenta
to Cherenkov-angle-based particle identification at higher momenta. Since our
analysis was performed in $\phi$ momentum bins, such behavior may introduce
a bias in the result if the kaon selection efficiency is taken as a constant value over 
the entire range. We therefore extract the kaon selection efficiency separately above
and below this momentum. We measure $\eps_{K^+}=(98.6\pm 1.2)\%$ and $\eps_{K^-}=(98.7\pm 1.2)\%$
for $p_\phi<1.2~\gevc$, and $\eps_{K^+}=(79.0\pm 1.8)\%$ and $\eps_{K^-}=(79.7\pm 1.9)\%$
for $p_\phi>1.2~\gevc$.

\begin{figure}[h]
\begin{center}
\includegraphics[width=3.5in]{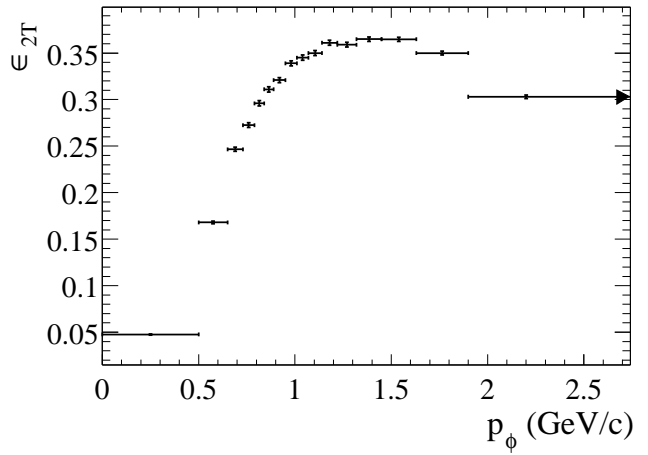}
   \caption{Efficiency $(\eps_{2T})$ as a function of $\phi$ momentum. The highest bin 
includes all entries above 1.9 \gevc. Errors shown are statistical only.}
   \label{fig:eff}
\end{center}
\end{figure}

The remaining efficiency to be estimated is $\eps_{2T}$, that of 
finding two charged tracks that originate from a $\phi$ meson, satisfy
the $R_2$ requirement, and have an invariant mass in the signal 
region (with no kaon identification requirement).
This efficiency is estimated from Monte Carlo.
Since the efficiency to reconstruct a $\phi$ depends on the $\phi$ 
momentum, differences between the momentum spectrum in data and the generated 
spectrum in the Monte Carlo sample must be considered. 
Therefore, the analysis was carried out separately in 16 bins of $\phi$ 
momentum. The bins are chosen to have equal (with the exception of 
the lowest momentum range) numbers of reconstructed $\phi$ mesons in the Monte Carlo.
Figure \ref{fig:eff} shows
the efficiency $\eps_{2T}$ as a function of the $\phi$ momentum.

\section{Results}
\label{sec:results}
The average multiplicity is calculated with the formula
\begin{equation}
{\cal B}(B\rightarrow \phi X) = 
{1\over 2N_{BB}\mathcal B(\phi\to K^+K^-)}
\sum_{i=1}^{16}{N^\phi_{B,i}\over \eps_i}~~~,
\label{multeq}
\end{equation}
where $N^\phi_{B,i}$ is the number of $\phi$ mesons in momentum bin $i$ found in the data
and assumed to come from $B$ mesons. This number is obtained by performing the background 
fit to the on-resonance data samples after subtracting the off-resonance 
data samples, scaled to the on-resonance luminosity.
The efficiency $\eps=\eps_{2T}\eps_{K^+}\eps_{K^-}$ is the 
product of the reconstruction efficiency and the kaon identification efficiencies
for each track. The quantity $N_{BB}$ is the number of \BB events in the data sample,
which is measured to be $N_{BB} = (18.7\pm 0.2)\times 10^6$ using a technique described
elsewhere \cite{bcount}. 

Since the analysis was performed in $\phi$ momentum bins, the efficiency in each bin has
very little dependence on the modeling of the $\phi$ spectrum, except for the
lowest-momentum bin, which includes the tracking detection limit.
We therefore sum the yield in the highest 15 bins and extrapolate the
result based on the simulated spectrum, so that the sum in
Equation~\ref{multeq} is replaced by

\begin{equation}
\sum_{i=1}^{16}{N^\phi_{B,i}\over \eps_i}\Longrightarrow
\sum_{i=2}^{16}{N^\phi_{B,i}\over \eps_i}\times
{\sum_{i=1}^{16}{N^\phi_{\mathrm{MC},i}}\over \sum_{i=2}^{16}{N^\phi_{\mathrm{MC},i}}}~~~.
\end{equation}

\noindent Here, $N^\phi_{\mathrm{MC},i}$ is the number of $\phi$ mesons
in the Monte Carlo sample in momentum bin $i$.

\begin{figure}[h]
\begin{center}
\includegraphics[width=3.5in]{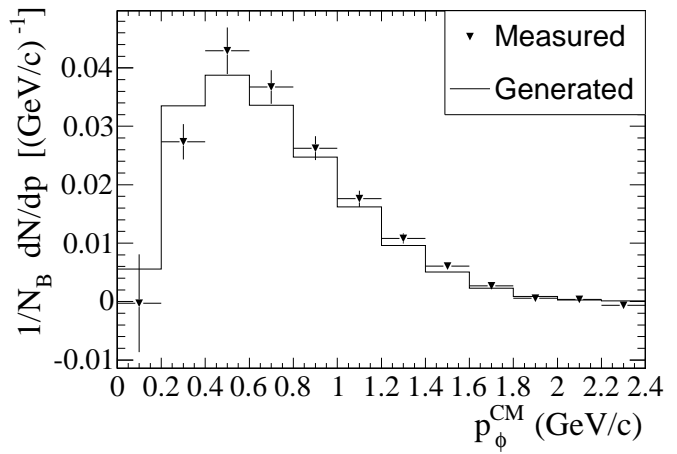}
\caption{Center-of-mass $\phi$ momentum spectra of the measured and 
Monte Carlo generated samples. The measured spectrum shows statistical and systematic errors combined, and the generated spectrum is normalized to
the measured multiplicity.
The lowest momentum bin has a negative central value due to the large 
continuum component but is consistent with positive values due to the large 
error.}
\label{fig:spectrum}
\end{center}
\end{figure}

Using ${\cal B}(\phi\to K^+K^-)=0.492 \pm0.006$ \cite{PDG},
we obtain ${\cal B}(B\rightarrow \phi X) = (3.41\pm0.06)\%$, where
the error is statistical only.
Figure \ref{fig:spectrum} shows the measured and 
simulated $\phi$ momentum spectra in the \FourS center-of-mass frame. We note
that the Monte Carlo sample predicts the observed $\phi$ momentum spectrum
reasonably well.

\section{Systematic uncertainties}
Systematic uncertainties are associated with all of the variables in 
Equation~1. Table~\ref{syst} lists the various systematic uncertainties,
which are described in detail below.

\begin{table}
\caption{Relative systematic uncertainties.}
\label{syst}
\begin{center}
\begin{tabular}{| l | l |} \hline
Source                        & $\Delta \mathcal{B} / \mathcal{B} (\%)$ \\ \hline
Combinatorial BG fitting      & {\bf 2.4} \\
On/Off scale factor           & {\bf 0.1} \\
$\mathcal{B}(\phi\to K^+K^-)$ & {\bf 1.2} \\
$N_{BB}$                      & {\bf 1.1} \\
$\eps_{2T}$ (total)       & {\bf 1.9}\\
\hspace{0.75 cm}$R_2$                          &\hspace{0.25 cm} 0.5 \\
\hspace{0.75 cm}Monte Carlo statistics                 &\hspace{0.25 cm} 0.3 \\
\hspace{0.75 cm}Monte Carlo $\phi$ modeling            &\hspace{0.25 cm} 0.7 \\
\hspace{0.75 cm}Tracking efficiency           &\hspace{0.25 cm} 1.6 \\ \hline
{\bf Total}                    & {\bf 3.4}\\ \hline
\end{tabular}
\end{center}
\end{table}

Two sources contribute to the uncertainty on $N^\phi_{Bi}$, the 
number of $\phi$ mesons from \BB events. One is the fitting procedure error 
which is described in Section \ref{section:bg} and was taken to be 2.4\%. 
The other is the error on the scaling factor (on/off-resonance) which contributes 
0.1\%  relative uncertainty on the measured average multiplicity.

The uncertainty on $\mathcal B(\phi\to K^+K^-)$ introduces a relative error of 1.2\%,
while that on $N_{BB}$ contributes 1.1\% relative uncertainty on the average multiplicity.

As $\eps_{2T}$ is obtained from Monte Carlo, differences 
between data and Monte Carlo give rise to systematic uncertainties. Several
components contribute to this systematic uncertainty:

\begin{itemize}
\item There is a systematic uncertainty related to the extrapolation of the 15-bin
yield to the full result, due to our limited knowledge of the $\phi$ spectrum
in $B$ decays. Since only about 60\% of $B$ meson decays are well 
understood (see Section~\ref{detector}), mismodeling of the remaining 40\% can affect the result.
To account for this model dependence, we study two Monte Carlo
subsamples representing extreme cases to make up the entire remaining 40\%.
The first subsample contains a $B$ meson undergoing a two-body
decay to a charm meson, with the charm meson undergoing a two-body decay 
to a $\phi$. In the second subsample, the $B$ meson 
undergoes a multi-body (greater than two) decay and the subsequent charm meson 
undergoes a multi-body decay into a final state that contains a $\phi$ meson.
These two cases yield very different kinematic distributions for the $\phi$ meson.
We measure the fraction of candidates in the lowest bin in each sample. 
We take the largest difference between these samples and the primary result
as a systematic uncertainty. It is found to be 0.7\%. The effect of $\phi$
polarization was similarly studied and found to have a negligible impact
on the result.

\item To establish the 
contribution to the systematic uncertainty from the simulation of $R_2$,
the number of $\phi$ mesons from $B\overline B$ events is estimated again 
without the $R_2$ requirement, and the same
procedure is applied to the Monte Carlo. The fraction of $\phi$ mesons from $B$ decays
with $R_2<0.25$ in data is (78.18\plm0.80)\% while this fraction in
Monte Carlo is (78.00\plm0.09)\%, in agreement within statistical errors.
We also investigate the two decay models mentioned above for their effect 
on the $R_2$ selection. We find that the largest difference between the models and our
Monte Carlo distribution is 0.5\%, and we take this difference as a systematic
uncertainty.

An additional test is performed by examining different continuum-suppression variables.
We study the angle between the $\phi$ direction in the center-of-mass frame and
thrust axis of the event, where the thrust was calculated both including and excluding the $\phi$
candidate. These two variables are each used in place of $R_2$ in order to suppress 
continuum events. We place appropriate criteria on these variables
to maintain similar efficiency to that of $R_2$ in our 
analysis. We then measure the efficiency of these requirements in data and Monte Carlo.
The ratio of efficiencies between data and Monte Carlo is found to
be $0.982\pm0.015$ for the first variable (with the $\phi$) and $1.007\pm0.015$ 
for the other.

\item Tracking performance is studied using control samples in data, and the
track-finding efficiency is found to be accurate to within 0.8\% per track. We
therefore assign a 1.6\% systematic uncertainty due to tracking efficiency.

\item Finally, the statistical uncertainty on $\eps_{2T}$ contributes a 0.3\%  
systematic uncertainty. 
\end{itemize}

The use of one single kaon selection efficiency for all $\phi$
   momenta was compared to the use of separate values above and below
   $p_{\phi}=1.2 \gevc$.  The observed difference in the average multiplicity
   was 0.9\%.
This is below the statistical error on the kaon identification efficiencies, hence no
additional error was assigned to this source. The statistical error
on the kaon selection efficiencies is treated as part of the statistical error
in this analysis as it is obtained from the same data set as our
signal and scales appropriately.

The above sources of systematic uncertainty are added in quadrature 
and yield a relative uncertainty on the average multiplicity of 3.4\%.

This analysis is repeated by replacing the 16 $p_\phi$ bins with 6 bins of 
$\theta_\phi$, the polar angle of the 
$\phi$ candidate with respect to the beam axis. As the total 
number of events is exactly the same as in the 
analysis described above, this is not an independent measurement, and can 
only serve to validate the fitting procedure.
The combinatorial background in these bins is significantly different in shape 
to that used in the primary analysis. The total yield of $\phi$ mesons from 
$B$ decays is found to differ by 0.88\% from the yield in the primary analysis ---
well within the assigned uncertainty for this source.

We also repeat the analysis using a different data set. We use a smaller
data set from the year 2000 in which the detector was operating under 
different conditions. This analysis yields $\mathcal B(B\to\phi X)=(3.34\pm 0.07)\%$
where the error is statistical only, entirely consistent with our primary result.

\section{Conclusion}

By selecting two identified oppositely-charged kaons from a sample of \FourS data
and subtracting the combinatorial and continuum background, we measure the
average multiplicity of $\phi$ mesons in $B$ meson decays.
Our measurement of $\mathcal B(B\to\phi X)=(3.41\pm0.06\pm0.12)\%$ is
consistent with both previous measurements at the $1.5\sigma$ level,
although it is significantly more precise.

We are grateful for the 
extraordinary contributions of our \pep2\ colleagues in
achieving the excellent luminosity and machine conditions
that have made this work possible.
The success of this project also relies critically on the 
expertise and dedication of the computing organizations that 
support \babar.
The collaborating institutions wish to thank 
SLAC for its support and the kind hospitality extended to them. 
This work is supported by the
US Department of Energy
and National Science Foundation, the
Natural Sciences and Engineering Research Council (Canada),
Institute of High Energy Physics (China), the
Commissariat \`a l'Energie Atomique and
Institut National de Physique Nucl\'eaire et de Physique des Particules
(France), the
Bundesministerium f\"ur Bildung und Forschung and
Deutsche Forschungsgemeinschaft
(Germany), the
Istituto Nazionale di Fisica Nucleare (Italy),
the Foundation for Fundamental Research on Matter (The Netherlands),
the Research Council of Norway, the
Ministry of Science and Technology of the Russian Federation, and the
Particle Physics and Astronomy Research Council (United Kingdom). 
Individuals have received support from 
the A. P. Sloan Foundation, 
the Research Corporation,
and the Alexander von Humboldt Foundation.

\end{document}